\documentclass[]{revtex4}
\begin{document}

\title{  Perturbative quantum gravity  in Batalin-Vilkovisky formalism}

\author{ Sudhaker Upadhyay\footnote {e-mail addresses: sudhakerupadhyay@gmail.com,\ \  
 sudhaker@boson.bose.res.in}}

\affiliation { S. N. Bose National Centre for Basic Sciences,\\
Block JD, Sector III, Salt Lake, Kolkata -700098, India. }
 
\begin{abstract}
In this Letter we consider the perturbative quantum gravity  on the super-manifold
which remains invariant under  absolutely  
anticommuting BRST and anti-BRST  transformations. 
In addition to that the
theory posses one more symmetry known as shift symmetry. 
The 
BRST invariant Batalin-Vilkovisky (BV) action for perturbative quantum gravity
is realized as a translation in Grassmann coordinate. Further
we  show  that the quantum master equation of the BV 
quantization method at one-loop order can
be translated to  have a superfield structure for the action. 
  However, the BRST as well as anti-BRST invariant BV action is constructed 
  in superspace with the help of two Grassmann coordinates.
  \end{abstract}
\maketitle

\section{  Introduction}	
The gauge theories are one of the basic ingredients in the search for a description of the fundamental interactions involving elementary particles.
Gauge invariance of such theories is
translated at the quantum level into the fermionic rigid
 BRST  invariance \cite{brst, tyu} and is important in the proof of unitarity, renormalizability and other aspects of field
theories \cite{brst,tyu,ht,wei}.
 The  perturbative quantum gravity as a gauge theory is a   subject of 
 current research 
with different aspects \cite{hata1, hata2, asch}.
 Such models of gravity were initially studied as attempts to unify gravity with electromagnetism \cite{ein}. These are   studied also due to their relevance in string theory \cite{ch,da,ah}.
 
 On the other hand BV formulation is known to be one of the most powerful method of quantization for different 
gauge field theories, supergravity theories and topological field theories in 
Lagrangian formulation \cite{ht,wei,bv,bv1,bv2,subm, bss, subm1}.
The BRST and the anti-BRST symmetries for perturbative quantum gravity
in four dimensional flat spacetime have been studied by many people
\cite{na,ku,ni} and their work has been summarized by N. Nakanishi and I. Ojima \cite{nn}.
The BRST symmetry in two dimensional curved spacetime has been thoroughly
studied \cite{yo,be,fr}. The BRST and the anti-BRST symmetries for topological quantum gravity in curved spacetime have also been studied \cite{ta,me}.
However, a superspace formalism for the BV action of different gauge theories has been studied recently \cite{subm,ba,fk,fai} and  we will generalize the  such results 
in the case of  perturbative quantum gravity in curved spacetime.

In this Letter we analyse the BRST as well as anti-BRST transformations for perturbative quantum 
gravity which are
absolutely anticommuting in nature. Further it is shown that the Lagrangian 
density for such theory
also remains invariant under shift symmetry transformations. The  BRST and anti-BRST 
transformations collectively with shift symmetry transformations are, known as
extended BRST and extended anti-BRST transformations respectively, constructed
extensively. 
In this formulation it is shown that the antifields in BV formulation get proper 
identification
naturally through using equations of motion
for auxiliary fields. The superspace formulation of  extended 
BRST invariant perturbative quantum theory is constructed  with help of one Grassmann coordinate.
The superspace description of quantum master equation at one-loop order is also
analysed. 
 However to develop the superspace 
formulation    for both extended BRST as well as 
anti-BRST invariant perturbative quantum theory we need two Grassmann coordinates.  

The plan of the Letter is as follows. In section II, we study the preliminaries 
about the perturbative quantum gravity.
The extended BRST 
invariant   theory of gravity 
is constructed in section III.
In section IV, we establish a  superspace formulation for such an extended BRST 
invariant theory.
The section V, is devoted to study the superspace description  of quantum master equation involved  BV technique. 
The extended anti-BRST transformation is explored in the  section VI. In section
VII, we 
construct the both extended BRST and anti-BRST 
transformations for such theory in superspace. In the last  section we discuss the 
results and draw the conclusions.

\section{ The perturbative quantum gravity: preliminarily}
We start with the Lagrangian density for pure gravity given by
\begin{equation} 
{\cal L}   =  \sqrt{ g}(R-2\lambda), \label{kin}
\end{equation}
where $\lambda$ is a cosmological constant and the  units are chosen  such that
  $16\pi G=1$. 
 
 In perturbative gravity one writes the full metric $g_{ab}^f$ in terms of a fixed
background metric $g_{ab}$ and small perturbations around it. 
The
small perturbation around the fixed background metric, denoted by  $h_{ab}$, is 
considered as a 
 field that is to be quantized. 
So, we can write 
\begin{equation}
g_{ab}^f=g_{ab}+h_{ab}.
\end{equation}
 The Lagrangian density given in Eq. (\ref{kin}) remains invariant under following 
 gauge transformation
 \begin{eqnarray}
 \delta_\Lambda h_{ab}=\nabla_a \Lambda_b +\nabla_b \Lambda_a + {\pounds}_{(\Lambda)} h_{ab},
 \end{eqnarray}
 where the Lie derivative for  $h_{ab}$ is given  by
 \begin{eqnarray}
 {\pounds}_{(\Lambda)} h_{ab}=\Lambda^c\nabla_c h_{ab} +h_{ac}\nabla_b \Lambda^c
 +h_{ cb}\nabla_a \Lambda^c,
 \end{eqnarray}
and therefore the theory for perturbative quantum gravity has some redundant degrees 
 of freedom.  These unphysical degrees of 
freedom give rise to constraints \cite{ht} in the canonical quantization and 
divergences in the partition function   in the path integral  quantization. In 
order to remove the
redundancy in degrees of freedom we need to fix the gauge by putting following  
gauge-fixing condition  \begin{equation}
G[h]_a=(\nabla^b h_{ab} -k\nabla_a h) =0,
\end{equation}
where $k\neq  1$. For $k=1$ the conjugate momentum corresponding to $h_{00}$ 
vanishes
and therefore the partition function diverges again. To ensure the unitarity a 
Faddeev-Popov  ghost term is also needed.
 
The  gauge-fixing term (${\cal L}_{gf} $) and ghost term ($ {\cal L}_{gh}$) 
corresponding to the above gauge-fixing condition are
given by
  \begin{eqnarray}
{\cal L}_{gf}&=&  \sqrt{ g}[ib^a(\nabla^b h_{ab}-k \nabla_a h)]  , \label{gfix}\\
{\cal L}_{gh}&=& i \sqrt{ g}\bar c^a \nabla^b [ \nabla_a c_b+ \nabla_b c_a- 
2kg_{ab}\nabla_c c^c +
  {\pounds}_{(c)} h_{ab}  -kg_{ab}g^{cd} {\pounds}_{(c)} h_{cd}], \nonumber\\
  &=&\sqrt{ g}\bar c^a M_{ab} c^b,
\end{eqnarray} 
where $
M_{ab}= i  \nabla_c [ \delta_b^c\nabla_a  + g_{ab}\nabla_c - 2k \delta_a^c\nabla_b 
+\nabla_b h^c _a -h_{ab}\nabla^c
-h^c_b\nabla_a 
-kg^c_ag^{ef}(\nabla_b h_{ef} +h_{eb}\nabla_f +h_{fb}\nabla_e)].
$

Now, the complete Lagrangian density is given by 
\begin{equation}
{\cal L}_C ={\cal L} +{\cal L}_{gf}+{\cal L}_{gh},  \label{com}
\end{equation}
which remains invariant under following BRST transformations,  
\begin{eqnarray}
s  h_{ab} = - (\nabla_a c_b +\nabla_b c_a +{\pounds}_{(c)} h_{ab}), \  
s c^a  =  -c_b\nabla^b c^a,     \ s  \bar c^a
= b^a,\ s  b^a  =  0.\label{sym}
\end{eqnarray}
This  Lagrangian density is also invariant under the following anti-BRST
transformations,  
\begin{eqnarray}
\bar s  h_{ab} = - (\nabla_a \bar c_b +\nabla_b \bar c_a +{\pounds}_{(\bar c)} 
h_{ab}), \  
\bar s \bar c^a  =  -\bar c_b\nabla^b \bar c^a,     \ \bar s    c^a
= -b^a,\ \bar s  b^a  =  0.\label{ansym}
\end{eqnarray}
It is easy to check that the above BRST and anti-BRST transformations are nilpotent as well as
absolutely anticommuting in nature i.e. 
\begin{eqnarray}
s^2=0,\ \bar s^2 =0,\ s\bar s +\bar s s =0.
\end{eqnarray} 
 
Now, we express the gauge-fixing and ghost part of the 
complete Lagrangian density as follows,
\begin{eqnarray}
{\cal L}_g &=& {\cal L}_{gf} +{\cal L}_{gh},\nonumber\\
&=&i s \sqrt{ g}[\bar c ^a (\nabla^b h_{ab} -k\nabla_a h)],\nonumber\\
&=&-i \bar s \sqrt{ g}[ c ^a (\nabla^b h_{ab} -k\nabla_a h)],\nonumber\\
&=&-\frac{1}{2}i s\bar s \sqrt{ g}  ( h^{ab} h_{ab}  ),\nonumber\\
&=&\frac{1}{2}i \bar s s\sqrt{ g}  ( h^{ab} h_{ab}  ).\label{g}
\end{eqnarray}
In the BV formalism, the gauge-fixing and ghost part of the Lagrangian density   
is generally expressed in terms of BRST variation of a gauge-fixed fermion.
It is straightforward to write the ${\cal L}_{g }$ given in Eq. (\ref{g})
in terms of gauge-fixed fermion $\Psi$ 
as    
\begin{equation}
{\cal L}_{g }=s  \Psi,
\end{equation}
where the expression for $\Psi$ is 
\begin{equation}
\Psi =i  \sqrt{ g}[\bar c ^a (\nabla^b h_{ab} -k\nabla_a h)].\label{gff}
\end{equation}
Now, we will analyse the extended BRST transformation for the perturbative
quantum gravity.
\section{  The BV formalism with  extended BRST  transformation }
In this section we study the extended BRST transformations for 
perturbative  quantum gravity in the BV context. To do so we extend the 
BV action by shifting the all fields as follows
\begin{equation}
\tilde{\cal L}_{g }= {\cal L}_{g} (h_{ab}-\tilde h_{ab}, c^a-\tilde c^a, \bar c^a 
-\tilde{\bar c}^a, b^a -\tilde b^a).
\end{equation}
 The above Lagrangian density remains invariant under following 
 extended BRST transformations,
\begin{eqnarray}
s  h_{ab} &=&\psi_{ab},\ s  \tilde h_{ab}= \psi_{ab}+ (\nabla_a c_b -\tilde 
\nabla_a \tilde c_b +\nabla_b c_a -\tilde \nabla_b \tilde c_a +{\pounds}_{(c)} h_{ab} -{\pounds}_{(\tilde c)}  \tilde h_{ab}),\nonumber\\  
s c^a  &=& \rho^a,\ s \tilde c^a=\rho^a +c_b\nabla^b c^a -\tilde c_b\tilde 
\nabla^b \tilde c^a,     \ s  \bar c^a
= B^a,\  s  \tilde{\bar c}^a =B^a-b^a +\tilde b^a,\nonumber\\
 s  b^a  &= &\chi^a,\ s \tilde b^a   = \chi^a, \ s [\psi_{ab},\rho^a, B^a, 
 \chi^a]= 0,\label{exsym}
\end{eqnarray}
where the fields $\psi_{ab}, \rho^a,  B^a$ and $ \chi^a$ are the 
associated ghost for the shift symmetric fields. Further we need 
to introduce the antighost fields $h_{ab}^\star, c_a^\star, \bar c_a^\star$ 
and $b_a^\star$ having opposite statistics to that of the respective
fields. The nilpotent BRST transformation for the antighost fields are
given by
\begin{eqnarray}
s h_{ab}^\star =-l_{ab},\ \ s c_a^\star =-m_a,\ \ s\bar c_a^\star =-n_a,\ \ s b_a^\star 
=-r_a,\ \ s[l_{ab}, m_a, n_a, r_a ]=0,\label{sym1}
\end{eqnarray} 
where $l_{ab}, m_a, n_a$ and $r_a$ are the Nakanishi-Lautrup type auxiliary fields.
 
 If we gauge fix the shift symmetry such that all the tilde fields vanish, then, of course, we   recover our original theory. To achieve this we choose the following shifted gauge-fixed Lagrangian density
\begin{eqnarray}
\tilde{\cal L}_{g } &= &- l_{ab}\tilde h^{ab} -h^{ab\star} (\psi_{ab}+ \nabla_a 
c_b -\tilde \nabla_a \tilde c_b +\nabla_b c_a -\tilde 
\nabla_b \tilde c_a +{\pounds}_{(c)} h_{ab} -{\pounds}_{(\tilde c)}  \tilde h_{ab} )  \nonumber\\
&+&m_a\tilde{\bar c}^a -c_a^\star (B^a-b^a +\tilde b^a)-n_a\tilde c^a +\bar 
c_a^\star (\rho^a +c_b\nabla^b c^a -\tilde c_b\tilde \nabla^b 
\tilde c^a)\nonumber\\
& +& r_a\tilde b^a+b_a^\star \chi^a, \label{la}
\end{eqnarray}
which remains invariant 
under the extended BRST symmetry transformations given in Eqs. (\ref{sym}) and 
(\ref{sym1}).

Now, it is straightforward to check  that using equations of motion of 
auxiliary fields, $l_{ab}, m_a, n_a, r_a$,   all the tilde fields  disappear from 
the above expression. The
 extended Lagrangian density $ \tilde {\cal L}_{g }$ then remains with the 
 following form:
\begin{eqnarray}
\tilde{\cal L}_{g } &= &   -h^{ab\star}(\psi_{ab}+ \nabla_a c_b  +\nabla_b c_a  
 +{\pounds}_{(c)} h_{ab} )  \nonumber\\
&-& c_a^\star (B^a-b^a  )  +\bar c_a^\star (\rho^a +c_b\nabla^b c^a)  +b_a^\star 
\chi^a. \label{per}
\end{eqnarray}
The gauge-fixed  fermion $\Psi$ depends only on the original fields, then a
general gauge-fixing Lagrangian density 
for perturbative quantum gravity with  original BRST symmetry 
will have the following form
\begin{eqnarray}
{\cal L}_{g} &= &s  \Psi =s  h_{ab}\frac{\delta\Psi}{\delta h_{ab}}+
s c_a \frac{\delta\Psi}{\delta c_a}+ s \bar{c}_a \frac{\delta\Psi}
{\delta\bar{c}_a} +
s b_a \frac{\delta\Psi}{\delta b_a} , \nonumber\\
&= &-\frac{\delta\Psi}{\delta h_{ab}}\psi_{ab}+
 \frac{\delta\Psi}{\delta c_a}\rho_a + \frac{\delta\Psi}
{\delta\bar{c}_a} B_a -
\frac{\delta\Psi}{\delta b_a} \chi_a.\label{psi}
\end{eqnarray}

Now, the total Lagrangian density ${\cal L}_{T} =  {\cal L}_0+
{\cal L}_{g }+\tilde{\cal L}_{g }$  is then given by
\begin{eqnarray}
{\cal L}_{T}&=&
  \sqrt{ g}(R-2\lambda) -h^{ab\star}(  \nabla_a c_b  +\nabla_b c_a  
 +{\pounds}_{(c)} h_{ab} )  
- c_a^\star b^a   +\bar c_a^\star ( c_b\nabla^b c^a) \nonumber\\ 
&-&\left(h_{ab}^\star +
\frac{\delta\Psi}{\delta h_{ab}}\right)\psi_{ab}+\left( \bar c_a^\star +
 \frac{\delta\Psi}{\delta c_a}\right)\rho_a -\left(c_a^\star -\frac{\delta\Psi}
{\delta\bar{c}_a}\right) B_a +\left (b_a^\star -
\frac{\delta\Psi}{\delta b_a} \right)\chi_a, 
\end{eqnarray}
where we have used the Eqs. (\ref{kin}), (\ref{per}) and (\ref{psi}).
Integration over
ghost fields  associated with the shift symmetry leads to the following 
identification  
\begin{eqnarray}
h_{ab}^\star =-
\frac{\delta\Psi}{\delta h_{ab}},\ \  \bar c_a^\star =-
 \frac{\delta\Psi}{\delta c_a},\ \
c_a^\star = \frac{\delta\Psi}
{\delta\bar{c}_a},\ \ b_a^\star = 
\frac{\delta\Psi}{\delta b_a}.
\end{eqnarray}
For the gauge-fixed fermion given in Eq. (\ref{gff}),
  the antifields associated with the theory are identified as 
\begin{eqnarray}
h_{ab}^\star & =& 
i\sqrt{g} (\nabla_b\bar c_a -kg_{ab}\nabla_c\bar c^c),\ \ \ \bar c_a^\star =0,
\nonumber\\
c_a^\star &=& i\sqrt{g} (\nabla^b h_{ab} -k \nabla_a h),\ \ \ b_a^\star =0. 
\label{antifield}
\end{eqnarray}
With these identification the total Lagrangian density reduces to the original 
theory for perturbative quantum gravity. 
Now, we are  able to write the  gauge-fixing part of the total Lagrangian density 
in terms of
the BRST variation of a generalized gauge-fixed fermion, i.e.
\begin{eqnarray}
 {\cal L}_{g} + \tilde{\cal L}_{g }&=&   s \left( h_{ab}^\star{\tilde 
 h}^{ab} -c_a^\star \tilde {\bar c}^a +\bar c_a^\star\tilde c^a -b_a^\star \tilde b^a
 \right).
\end{eqnarray}
  As 
expected the ghost number of   ($  h_{ab}^\star{\tilde 
 h}^{ab}-c_a^\star \tilde {\bar c}^a +\bar c_a^\star\tilde c^a -b_a^\star \tilde b^a
$)   is equal to $-1$.  
In the next section, we construct the superspace formulation of   extended BRST invariant
perturbative quantum gravity
theory. 
\section{  Extended BRST invariant superspace formulation at classical level}    
In Ref.  \cite{subm}, a superspace formulation for the shifted
field approach to the Batalin-Vilkovisky action at the classical 
level was presented for the case of the higher form gauge
theory. Here we will  review this formulation, presenting it in a general way for the perturbative quantum gravity.

To write a superspace
formalism of the extended BRST invariant theory  we consider a superspace with 
coordinates $(x^a, \theta)$ where  $\theta$ is fermionic coordinate. 
In such a superspace the ``superconnection" $2$-form can be written as  
\begin{equation}
\omega^{(2)} =\frac{1}{2 !}{\cal H}_{ab}(x, \theta ) (dx^a\wedge dx^b ) +
 {\cal C}_{a}(x, \theta ) (dx^a\wedge d\theta ), 
\end{equation}
where $d $ is an exterior derivative and is defined as $d=dx^a\nabla_a 
+d\theta \partial_\theta$.
The requirement for super curvature (field
strength   $F^{(3)}= d\omega^{(2)}$) to vanish
along the $\theta$ direction restricts 
the component of the superfields  to have   following form
\begin{eqnarray}
{\cal H}_{ab}(x, \theta ) &=& h_{ab} (x) +\theta   (sh_{ab} )= h_{ab} (x) +\theta  
\psi_{ab},\nonumber\\
{\cal C}_{a}(x, \theta ) &=& c_a (x) +\theta  (s c_a )= c_a (x) +\theta \rho_a.
\end{eqnarray}
In this formalism, the antighosts and
auxiliary field  have to be introduced as additional superfields of the form 
\begin{eqnarray}
 {\bar {\cal C}_a} (x, \theta ) &=& \bar c_a (x) +\theta  (s \bar c_a )=\bar c_a 
 (x) +\theta  B_a ,\nonumber\\
{\cal B}_{a}(x, \theta ) &=& b_a (x) +\theta  (s b_a )=  b_a (x) +\theta \chi_a. 
\end{eqnarray}
Similarly, we  define all the superfields corresponding to the shifted fields  involved in extended  action  as 
\begin{eqnarray}
\tilde{\cal H}_{ab}(x, \theta ) &=&\tilde h_{ab} (x) +\theta   (s\tilde h_{ab} )= 
\tilde h_{ab} (x) +\theta  (\psi_{ab}+ \nabla_a c_b -\tilde \nabla_a \tilde c_b 
 \nonumber\\
&+&\nabla_b c_a -\tilde \nabla_b \tilde c_a+{\pounds}_{(c)} h_{ab} -
{\pounds}_{(\tilde c)}  \tilde h_{ab}),\nonumber\\
\tilde{\cal C}_{a}(x, \theta ) &=&\tilde c_a (x) +\theta  (s\tilde c_a )=\tilde 
c_a (x) +\theta ( \rho_a +c_b\nabla^b c_a -\tilde c_b\tilde \nabla^b \tilde c_a),
\nonumber\\
\tilde {\bar {\cal C}_a} (x, \theta ) &=& \tilde{\bar c}_a (x) +\theta  (s 
\tilde{\bar c}_a )=\tilde{\bar c}_a (x)
 +\theta  ( B_a-b_a +\tilde b_a),\nonumber\\
\tilde{\cal B}_{a}(x, \theta ) &=& \tilde b_a (x) +\theta ( s\tilde b_a )= \tilde 
b_a (x) +\theta \chi_a. 
\label{sufi}
\end{eqnarray} 
In addition, the super antifields   have the following form
\begin{eqnarray}
 {\cal H}^\star_{ab}(x, \theta ) &=& h^\star_{ab} (x) +\theta  ( s h^\star_{ab} )=   
 h^\star_{ab} (x) -\theta  l_{ab},\nonumber\\
 {\cal C}^\star_{a}(x, \theta ) &=&  c^\star_a (x) +\theta (s  c^\star_a )=  
 c^\star_a (x) -\theta  m_a,\nonumber\\
 {\bar {\cal C}^\star_a} (x, \theta ) &=&  {\bar c}^\star_a (x) +\theta ( s {\bar 
 c}^\star_a )={\bar c}^\star_a (x)
 -\theta n_a,\nonumber\\
 {\cal B}^\star_{a}(x, \theta ) &=&  b^\star_a (x) +\theta ( s\tilde b_a )=  
 b^\star_a (x) -\theta r_a.\label{antisufi}
\end{eqnarray}
Using  Eqs. (\ref{sufi}) and (\ref{antisufi}) we are able to write
the following expressions    
\begin{eqnarray}
 \frac{\delta}{\delta\theta} {\cal H}^\star_{ab}\tilde{\cal H}^{ab}
 &=&  -l_{ab}\psi^{ab}-h^{ab\star}  (\psi_{ab}+ \nabla_a c_b -\tilde \nabla_a 
 \tilde c_b +\nabla_b c_a -\tilde \nabla_b \tilde c_a \nonumber\\
&+&{\pounds}_{(c)} h_{ab} -{\pounds}_{(\tilde c)}  \tilde h_{ab}),
\nonumber\\
 \frac{\delta}{\delta\theta}\bar{\cal C}_{a}^\star  \tilde{\cal C}^{a}
 &=&  - n_a\tilde c^a+\bar c_a^\star  (\rho^a +c_b\nabla^b c^a -\tilde c_b\tilde 
 \nabla^b \tilde c^a ),
\nonumber\\
- \frac{\delta}{\delta\theta}{\cal C}_{a}^\star  \tilde{\bar{\cal C}}^{a}
 &=&    m_a\tilde {\bar c}^a- c_a^\star  (B^a -b^a +\tilde b^a ),
\nonumber\\
 -\frac{\delta}{\delta\theta} {\cal B}_a^\star  \tilde{\cal B}^a
 &=&  r_a\tilde b^a  +b_a^\star \chi^a.
 \end{eqnarray} 
Here we notice that the gauge-fixed   
Lagrangian density for shift symmetry  given in Eq. (\ref{la}) can be written  in 
the superspace formulation as
\begin{eqnarray}
\tilde{\cal L}_{g}= \frac{\delta}{\delta\theta}\left[{\cal 
H}^\star_{ab}\tilde{\cal H}^{ab}+\bar{\cal C}_{a}^\star  \tilde{\cal C}^{a}-
{\cal C}_{a}^\star  \tilde{\bar{\cal C}}^{a}-{\cal B}_a^\star  \tilde{\cal 
B}^a\right].\label{super}
\end{eqnarray}
Being the  $\theta$ component of 
superfields the extended Lagrangian density $\tilde{\cal L}_{g}$ remains invariant under the extended BRST transformation. If the 
gauge-fixing fermion depends only on the original fields, then one can define the 
fermionic superfield  $\Gamma $ as
\begin{eqnarray}
{\Gamma  } &=&\Psi +\theta s  \Psi,\nonumber\\ 
  &=&\Psi  + \theta \left[-\frac{\delta\Psi}{\delta h_{ab}}\psi_{ab}+
 \frac{\delta\Psi}{\delta c_a}\rho_a + \frac{\delta\Psi}
{\delta\bar{c}_a} B_a -
\frac{\delta\Psi}{\delta b_a} \chi_a \right].
\end{eqnarray}
With these realization the original gauge-fixing Lagrangian density ${\cal L}_{g } 
$ in the superspace
formalism can be expressed as 
\begin{equation}
{\cal L}_{g } =\frac{\delta\Gamma }{\delta\theta}.\label{lagr}
\end{equation}
Further we notice that invariance  of ${\cal L}_{g }$ under the extended BRST 
transformation is assured as it is
 $\theta$ component of super gauge-fixing fermion.  
\section{ The  quantum master equation of perturbative quantum gravity in 
 superspace} 
In this section we  first investigate the BRST variation of the quantum action in 
the standard BV quantization
method. Then we will analyse the corresponding behavior in the
superspace approach with one Grassmannian coordinate. For this purpose we first define the
vacuum functional for perturbative quantum gravity
with ordinary field in BV formulation as
\begin{equation}
Z_\Psi = \int \prod D\Phi \exp \left[\frac{i}{\hbar} W\left( \Phi, \Phi^\star = 
\frac{\partial \Psi}{\partial \Phi}\right) \right], 
\end{equation}
where $\Phi$ and  $\Phi^\star$ are the generic notation for all fields
and corresponding antifields involved in the 
the theory. However, $W$ is the
 extended action corresponding to the Lagrangian density given in Eq. (\ref{com}).
 
 The condition of gauge independence of generating functional
 is translated into the so-called quantum master equation given by
 \begin{equation}
 \frac{1}{2}(W, W)= i\hbar \Delta W,\label{bra}
 \end{equation}
 where the antibracket $(F, G)$ and operator $\Delta$ are
 defined as
 \begin{eqnarray}
 (X, Y) &=&\frac{\partial_r F}{\partial \Phi}\frac{\partial_l G}{\partial
  \Phi^\star}-\frac{\partial_r F}{\partial \Phi^\star}\frac{\partial_l G}{\partial
  \Phi },\nonumber\\
  \Delta &=& \frac{\partial_r }{\partial\Phi} \frac{\partial_l }
  {\partial\Phi^\star}.
  \end{eqnarray}
   The quantum action can be extended upto the one-loop order correction
   as
   \begin{equation}
   W(\Phi, \Phi^\star )=S_C(\Phi, \Phi^\star ) +\hbar M_1(\Phi, \Phi^\star ),
   \end{equation}
where $S_C$ is the complete action for Lagrangian density given in Eq. (\ref{com}) and $M_1$ appears from nontrivial measure factors.    
     
   The behavior of $W$ with respect to BRST 
   transformations can be given by\cite{ht}
   \begin{equation}
   s W =i\hbar\Delta W.
   \end{equation}
   For non-anomalous gauge theory upto first-order correction $M_1$ 
   does not depend on antifields. 
   In this situation, the BRST transformations of the complete 
   action $S_C$ and $M_1$
   are given by
   \begin{equation}
   s S_C=0, \ \ s M_1 =i\Delta S_C.
   \end{equation}
    Now we apply the $\Delta$ operator on total action (along with shifted
    fields) as
    \begin{equation}
    \Delta S_C=\Delta S_T =\frac{\partial_r }{\partial\Phi} \frac{\partial_l }{\partial\Phi^\star}S_T,
    \end{equation}
here $\Phi$ and $\Phi^\star$ includes all the fields, shifted fields,
 ghosts and   and corresponding antighosts fields. 
   Therefore, at one-loop order, we must build up a superfield
   \begin{equation}
   {\cal M}_1 = M_1 +\theta i\Delta S_T.
   \end{equation}
However, the extended action in the superspace will have following form
\begin{equation}
{\cal W} =W +\theta i\hbar \Delta W.
\end{equation}
The $\Delta$ operator in the superspace with one Grassmann coordinate
can be defined as
\begin{eqnarray}
   \tilde \Delta = \frac{\partial_r }{\partial\Phi (x,\theta)} \frac{\partial_l }
  {\partial\Phi^\star(x,\theta )}.
  \end{eqnarray}
  The quantum master equation in superspace is simply described 
  by 
  \begin{eqnarray}
  \frac{\partial}{\partial \theta }{\cal W}=i\hbar \tilde \Delta {\cal W}.
  \end{eqnarray}
 Here  we notice that  the role of generator
of BRST transformations is essentially played by the differentiation with 
respect to $\theta$. Therefore, enlarging the configuration space with the variable $\theta$, we are equipping it with
Grassmannian translations that reproduce the effect of the antibrackets
given in Eq. (\ref{bra}).
 
\section{  Extended anti-BRST transformation} 
 Let us next generalize the anti-BRST transformation given in Eq. (\ref{ansym}).
 For that purpose we construct the extended BRST transformation under which the 
 extended Lagrangian density remains invariant as follows
 \begin{eqnarray}
\bar s \tilde h_{ab} &=&h^\star_{ab},\ \bar s   h_{ab}= h^\star_{ab}- (\nabla_a 
\bar c_b -\tilde \nabla_a \tilde {\bar c}_b +\nabla_b 
\bar c_a -\tilde \nabla_b \tilde {\bar c}_a +{\pounds}_{(\bar  c)} h_{ab} 
-{\pounds}_{(\tilde {\bar c})}  \tilde h_{ab}),\nonumber\\  
\bar s\tilde c^a  &=& c^{a\star},\ \bar s  c^a=c^{a\star} -b^a +\tilde b^a,     
\ \bar s  \tilde{\bar c}^a
= \bar c^{a\star},\  \bar s  {\bar c}^a = \bar c^{a\star} -\bar c_b\nabla^b \bar 
c^a +\tilde {\bar c}_b\tilde \nabla^b \tilde {\bar c}^a,
\nonumber\\
\bar s \tilde b^a  &=& b^{a\star},\ \bar s  b^a   =  b^{a\star}, \ \bar s \
[h^\star_{ab}, c^{a\star}, \bar c^{a\star}, b^{a\star}]= 0.
\label{ab} 
\end{eqnarray} 
We note that the above anti-BRST transformation   absolutely anticommute   with 
the extended
BRST transformation given in Eq. (\ref{exsym}), i.e. $\{s, \bar s \}  =0$.

The anti-gauge-fixing  fermion   $\bar\Psi$ (gauge-fixing fermion in case of anti-
BRST transformation)
 for such theory is defined as
\begin{equation}
\bar\Psi =-i\sqrt{g}c^a(\nabla^b h_{ab}-k\nabla_a h).
\end{equation}
The  gauge-fixing as well as the ghost part of the Lagrangian density  can be written in terms of anti-BRST variation of $\bar\Psi$ as
\begin{equation}
{\cal L}_{g}=\bar s  \bar\Psi. 
\end{equation}
 
The ghost fields associated with the shift symmetry have the following extended
anti-BRST transformations,
\begin{eqnarray}
\bar s \psi_{ab}&=& l_{ab},\ \ \bar s \rho_{a} = m_a,\ \
\bar s B_{a} =n_{a},\ \
\bar s \chi_a = r_a,\nonumber\\
 \bar s   l_{ab} &=& 0,\ 
\bar s  m_a = 0,\ 
\bar s n_a =  0,\   \bar s r_a =  0.\label{exan}
\end{eqnarray}
The  transformations  given  in Eqs. (\ref{ab}) and (\ref{exan}) are the
extended anti-BRST transformations under which the Lagrangian density 
with shifted fields remains invariant.
 
\section{  Extended BRST and anti-BRST invariant superspace formulation}
The extended
BRST  and   anti-BRST invariant Lagrangian density is written  in superspace   with the help of  two Grassmannian 
coordinates $\theta$ and $\bar\theta$.
All the superfields in this superspace are the function of coordinates $(x^a, \theta, 
\bar\theta)$. The 
``super connection" $2$-form ($\omega^{(2)}$) 
is defined in this case as
\begin{eqnarray}
\omega^{(2)} =\frac{1}{2 !}{\cal H}_{ab}(x^a, \theta, \bar\theta ) (dx^a\wedge 
dx^b ) +
 {\cal C}_{a}(x, \theta, \bar\theta ) (dx^a\wedge d\theta ) + \bar{\cal{ C}}_a (x, 
 \theta, \bar\theta ) 
(dx^a\wedge d\bar\theta ).
\end{eqnarray} 
Here the exterior derivative has the following form $d=dx^a\nabla_a 
+d\theta\nabla_\theta +
d\bar \theta\nabla_{\bar\theta}$.
 
Requiring the field strength to vanish along all extra directions
 $\theta$ and $\bar\theta$ determines the superfields to have the following forms 
\begin{eqnarray}
 {\cal H}_{ab}(x^a, \theta, \bar\theta ) &=& h_{ab} (x^a) +\theta \psi_{ab}
+\bar\theta [ h_{ab}^\star -(\nabla_a\bar c_b-\tilde\nabla_a\tilde{\bar 
c}_b+\nabla_b\bar c_a-\tilde\nabla_b\tilde{\bar c}_a\nonumber\\
&+&
{\pounds}_{(\bar c)}h_{ab}-{\pounds}_{\tilde {(\bar c)}}\tilde h_{ab}
  )] +\theta\bar\theta [l_{ab} -\nabla_a b_b +\tilde\nabla_a\tilde b_b  +\nabla_b 
 b_a\nonumber\\
 &-&\tilde\nabla_b\tilde b_a  
 -s({\pounds}_{(\bar c)}h_{ab})+s({\pounds}_{({\tilde {\bar c}})}\tilde h_{ab} )],
 \nonumber\\
 \tilde{\cal H}_{ab}(x^a, \theta, \bar\theta ) &=&\tilde h_{ab} (x^a) + \theta 
 [\psi_{ab} +(\nabla_a  c_b-\tilde\nabla_a\tilde{  c}_b+\nabla_b 
 c_a-\tilde\nabla_b\tilde{ c}_a \nonumber\\
 &+ &
{\pounds}_{(  c)}h_{ab}-{\pounds}_{{( \tilde  c)}}\tilde h_{ab}
  )] +\bar \theta h^\star_{ab}
+\theta\bar\theta  l_{ab},\nonumber\\
{\cal C}_{a }(x^a, \theta, \bar\theta ) &=&c_a (x^a) +\theta\rho_a +\bar \theta (
c_a^\star -b_a+\tilde b_a)+\theta\bar\theta m_a,
\nonumber\\
\tilde{\cal C}_{a }(x^a, \theta, \bar\theta ) &=&\tilde c_a (x^a) +\theta[\rho_a 
+c_b\nabla^b c_a -
\tilde c_b\tilde\nabla^b \tilde c_a ]+\bar \theta c_a^\star+\theta\bar\theta m_a,
\nonumber\\
\bar{\cal C}_{a }(x^a, \theta, \bar\theta ) &=&\bar c_a (x^a) +\theta B_a +\bar 
\theta (\bar c_a^\star -\bar c_b\nabla^b \bar c_a
+\tilde{\bar c}_b\tilde\nabla^b\tilde{\bar c}_a)+\theta\bar\theta [n_a -
B_b\nabla^b \bar c_a\nonumber\\
 &+ &
\bar c_b\nabla^b B_a+(B_b-b_b +\tilde b_b )\nabla^b \tilde{\bar c}_a -\tilde{\bar c}_b \nabla^b  
(B_a-b_a +\tilde b_a )],
\nonumber\\
\tilde{\bar{\cal C}}_{a }(x^a, \theta, \bar\theta ) &=&\tilde {\bar c}_a (x^a) 
+\theta[B_a-b_a +\tilde b_a ]+\bar \theta \bar 
c_a^\star+\theta\bar\theta n_a,
\nonumber\\
 {\cal B}_{a }(x^a, \theta, \bar\theta ) &=&b_a (x^a) +\theta \chi_a +\bar \theta 
 b^\star_a +\theta\bar\theta r_a,
\nonumber\\
 \tilde{\cal B}_{a }(x^a, \theta, \bar\theta ) &=&\tilde b_a (x^a) +\theta \chi_a 
 +\bar \theta b^\star_a +\theta\bar\theta r_a.
\end{eqnarray}
Using the above expressions the Lagrangian density $\tilde {\cal L}_{g }$ given in Eq. (\ref{la}) can be written   as 
\begin{eqnarray}
\tilde {\cal L}_{g } 
=-\frac{1}{2}\frac{\delta}{\delta\bar\theta}\frac{\delta}{\delta\theta}\left[\tilde 
{\cal H}_{ab} \tilde {\cal H}^{ab} + {\tilde {\bar{\cal C}}}_a \tilde{\cal C}^a 
-{\tilde {\cal C}}_a \tilde{\bar{\cal C}}^a -\tilde{\cal B}_{a } \tilde {\cal 
B}^{a } \right].\label{barl}
\end{eqnarray}
Being the   $\theta\bar\theta$ component of the 
superfields   the $\tilde {\cal L}_{g } $ remains invariant under 
extended BRST as well as
extended anti-BRST transformations.
Furthermore we define the super gauge-fixed fermion  as
\begin{equation}
\Gamma  (x, \theta, \bar\theta )=\Psi +\theta (s \Psi )+\bar\theta (\bar s \Psi) 
+\theta
\bar\theta (s \bar s \Psi),
\end{equation} 
to express the ${\cal L}_{g }$ as $\frac{\delta}{\delta\theta}\left[\Gamma  (x, 
\theta, 
\bar\theta) \right]$. 
The  $\theta\bar\theta$ component of $\Gamma  (x, \theta, \bar\theta )$  vanishes 
due to equations of motion
in the theories having both BRST and anti-BRST invariance. 
 
The complete  gauge-fixing and ghost part of the Lagrangian density  which is 
invariant under both 
extended BRST and extended anti-BRST 
transformations  can therefore be written as
\begin{eqnarray} 
{\cal L}_{g }+\tilde {\cal L}_{g }
=-\frac{1}{2}\frac{\delta}{\delta\bar\theta}\frac{\delta}{\delta\theta}\left[
\tilde {\cal H}_{ab} \tilde {\cal H}^{ab} + {\tilde {\bar{\cal C}}}_a \tilde{\cal 
C}^a -{\tilde {\cal C}}_a \tilde{\bar{\cal C}}^a -\tilde{\cal B}_{a } \tilde {\cal 
B}^{a } \right] 
+\frac{\delta}{\delta\theta}\left[
\Gamma (x, \theta, \bar\theta)\right]. 
\end{eqnarray}
 Using equations of motion for auxiliary fields and the ghost fields associated 
 with shift symmetry
the antifields can be calculated. With these antifields the original gauge-fixed 
Lagrangian density in BV 
formulation can be recovered.   

Here we note that the
 one can generalize the quantum master equation
  of BV formalism in superspace for two Grassmannian coordinates
  following the section V.
\section{  Conclusions}
Although the superspace formulation for gauge theory have been done, we have 
generalized it for  perturbative quantum gravity.
The
BV formulation represents a very powerful
framework for the quantization of the most general 
gauge theories where the gauge algebra
is open or closed.
We have considered the perturbative quantum gravity   
in such framework. In such formalism one extends the configuration space by 
introducing antifields corresponding to all the original fields present in the 
theory. These antifields get identification
with the functional derivative of a gauge fixing fermion with respect to the 
corresponding fields.
We  have analysed the extended BRST and anti-BRST transformations
 (including the shift symmetry transformations) for this theory where we  have 
 shown that the antifields 
get its  proper identification automatically using equations of motion of 
auxiliary fields. Further, it has been shown that
to write the superspace formulation of extended BRST invariant
theory one needs one fermionic coordinate.
 We have shown that the  master equation of the BV formalism can be represented as 
 the requirement of a superspace structure for the quantum action.
 The quantum master equation at one-loop order has realized as a  translation 
 in $\theta$ variable.
  However, for both extended BRST and anti-BRST 
 invariant theory the superspace
description need   two Grassmannian coordinates.
In BV formalism antifields dependent terms (quantum corrections) of master equation 
must satisfy the consistency conditions proposed by Wess-Zumino as the theory becomes anomalous. 
The study of anomalies with antifields included in the action is based on the Zinn-Justin version \cite{zin}
of BV master equation.
It will be interesting to analyse the perturbative  theory 
of quantum gravity in superspace for higher order in $\hbar$
where the  quantum corrections depend on antifields.

\section*{Acknowledgments}
I thankfully acknowledge Dr. Mir Faizal for his kind discussions on the subject.

\end{document}